# Enhanced Tiny Encryption Algorithm with Embedding (ETEA)

Dr. Deepali Virmani, Nidhi Beniwal, Gargi Mandal, Saloni Talwar
BPIT, Rohini, Sector-17,New Delhi
deepalivirmani@gmail.com
BPIT, Rohini, Sector-17,New Delhi
nidhibeniwal1991@gmail.com
BPIT, Rohini, Sector-17,New Delhi
gargimandal98@gmail.com
BPIT, Rohini, Sector-17,New Delhi
salonitalwar09@gmail.com

## Abstract

As computer systems become more pervasive and complex, security is increasingly important. Secure Transmission refers to the transfer of data such as confidential or proprietary information over a secure channel. Many secure transmission methods require a type of encryption. Secure transmissions are put in place to prevent attacks such as ARP spoofing and general data loss. Hence, in order to provide a better security mechanism, in this paper we propose Enhanced Tiny Encryption Algorithm with Embedding (ETEA), a data hiding technique called steganography along with the technique of encryption (Cryptography). The advantage of ETEA is that it incorporates cryptography and steganography. The advantage proposed algorithm is that it hides the messages.

## Indexing terms

Cryptography, Steganography, Secure Transmission

## INTRODUCTION

The rapid growth and widespread use of electronic data processing and electronic business conducted through the Internet, along with numerous occurrences of international terrorism, fueled the need for better methods of protecting the computers and the information they store, process and transmit. Information is the primary commodity in today's world. As technology advances, the need to protect information to ensure its confidentiality, integrity, and availability to those whom need it for making critical personal, business, or government decisions becomes more important. The advancement of technology, the Internet, and information sharing has had both positive and negative impacts. One of the negative impacts was the large increase in new "information" threats. These threats have raised a number of concerns about how information is secured and maintained.

As computer systems become more pervasive and complex, security is increasingly important. Cryptographic algorithms and protocols constitute the central component of systems that protect network transmissions and store data. The security of such systems greatly depends on the methods used to manage, establish, and distribute the keys employed by the cryptographic techniques. Hence, in order to provide a better security mechanism, in this paper we propose Enhanced Tiny Encryption Algorithm with Embedding (ETEA), a data hiding technique called steganography [7] along with the technique of encryption (Cryptography). This paper is mainly proposed for providing security during transmission of data across the network. In this the sender encrypts the data into some form and embed it inside a Video by using "Enhanced Tiny Encryption Algorithm". This algorithm has been used because it requires less memory. It uses only simple operations, therefore it is easy to implement. The Tiny Encryption Algorithm (TEA) is a block cipher encryption algorithm that is very simple to implement, has fast execution time, and takes minimal storage space. By using the Input/Output packages the Steganography will read the video file and encrypted data and takes whole it as a video file. Thus, ETEA is a combination of Cryptography(TEA) and Steganography (Input/Output packages) together. So whenever the hacker tries to open the file, only video file is visible to them. Then this video file is sent to the network. File is sent by using Client-server technology, Threads, Ipv4 addressing.

# 1. LITERATURE REVIEW

## 1.1 Cryptography

Cryptography is the practice and study of techniques for secure communication in the presence of third parties (called adversaries) [1]. More generally, it is about constructing and analyzing protocols that overcome the influence of adversaries and which are related to various aspects in information security such as data confidentiality, data integrity, authentication, and non-repudiation. Modern cryptography intersects the disciplines of mathematics, computer science, and electrical engineering. Applications of cryptography include ATM cards, computer passwords, and electronic commerce [7] [10].

Cryptography prior to the modern age was effectively synonymous with encryption, the conversion of information from a readable state to apparent nonsense. The originator of an encrypted message shared the decoding technique needed to recover the original information only with intended recipients, thereby precluding unwanted persons to do the same. Since World War I and the advent of the computer, the methods used to carry out cryptology have become increasingly complex and its application more widespread.

## 1.2 Steganography

Steganography is an art and science of hiding information within other information. The word itself comes from Greek and means hidden writing. In recent years cryptography become very popular science. As steganography has very close to cryptography and its applications, we can with advantage highlight the main differences. Cryptography is about concealing the content of the message. At the same time encrypted data package is itself evidence of the existence of valuable information. Steganography goes a step further and makes the ciphertext invisible to unauthorized users. Hereby we can define steganography as cryptography with the additional property that its output looks unobtrusively [9][8].

## 1.3 TEA

The Tiny Encryption Algorithm (TEA) is a block cipher encryption algorithm that is very simple to implement, has fast execution time, and takes minimal storage space. The included example is to be compiled and used on a LabVIEW FPGA target [2] [3].

In cryptography, the Tiny Encryption Algorithm (TEA) is a block cipher notable for its simplicity of description and implementation, typically a few lines of code. TEA is very secure. There have been no known successful cryptanalyses of TEA. It's believed to be as secure as the IDEA algorithm, designed by Massey and Xuejia Lai. It uses the same mixed algebraic groups technique as IDEA, but it's very much simpler, hence faster [10].

The minor weaknesses identified by David Wagner at Berkeley are unlikely to have any impact in the real world, and one can always implement the new variant TEA which addresses them.

The mixing portion of TEA seems unbroken but related key attacks are possible even though the construction of 232 texts under two related keys seems impractical is one of the weaknesses. The second weakness is that the effective length of the keys is 126 bits not 128 does affect certain potential applications but not the simple cypher decypher mode.

## 1.4 Embedding inside Video

Data which hold effective information often has some redundancy. End users usually tend to think that redundancy is evil which cost extra money, as more disk space or network bandwidth is needed. Well, they are partially right, but optimal compression hardly ever exists. Moreover common compress ratio is mostly question of efficiency. Now we know, there are almost always few bytes, one can play with, without destroying carried information. When information is hidden inside video the program or person hiding the information will usually use the DCT (Discrete Cosine Transform) method.

DCT works by slightly changing the each of the images in the video, only so much though so it's isn't noticeable by the human eye. To be more precise about how DCT works, DCT alters values of certain parts of the images, it usually rounds them up. For example if part of an image has a value of 6.667 it will round it up to 7.Steganography in Videos [9] is similar to that of Steganography in Images, apart from information is hidden in each frame of video. When only a small amount of information is hidden inside of video it generally isn't noticeable at all, however the more information that is hidden the more noticeable it will become.

## 2. ETEA: PROPOSED WORK

In this paper, we have proposed Enhanced TEA(Tiny Encryption Algorithm) with embedding(ETEA) using input output packages of Java [5][2][4]. Enhanced TEA provides security during transmission of data across the network. In TEA, only encryption was possible. But in our proposed algorithm ETEA, encryption and embedding both are combined to provide high level of security to the data so that it couldn't be hacked.

The encrypted data using TEA is embedded in a Video file using Steganography and Input/Output Packages. This file can be transferred through network with high security to another user. Receiver can de-embed the video file and decrypt the original data using same key used at the time of encryption.

The following notations are necessary here:

- Hexadecimal numbers will be subscripted as "h," e.g., 10 = 16. h
- Bitwise Shifts: The logical left shift of x by y bits is denoted by x << y. The logical right shift of x by y bits is denoted by x >> y.
- Bitwise Rotations: A left rotation of x by y bits is denoted by x <<< y. A right rotation of x by y bits is denoted by x >>> y.
- Exclusive-OR: The operation of addition of n-tuples over the field (also known as 2F exclusive-or) is denoted by x⊕y.

The Enhanced Tiny Encryption Algorithm is a Feistel type cipher that uses operations from mixed algebraic groups. A dual shift causes all bits of the data and key to be mixed repeatedly [6].

The key schedule algorithm is simple; the 128-bit key K is split into four blocks of 32-bits each K = ( K[0], K[1], K[2], K[3]). In a Feistel cipher, the text being encrypted is split into two halves. The round function, F, is applied to one half using a sub key and the output of F is (exclusive-or-ed (XORed)) with the other half. The two halves are then swapped. Each round follows the same pattern except for the last round where there is often no swap.

- Each round i has inputs Left[i-1] and Right[i-1], derived from the previous round, as well as a sub key K[i] derived from the 128 bit overall K.

- The sub keys K[i] are different from K and from each other.

- The constant delta =(51/2-1)*231 =9E3779B h , is derived from the golden number ratio to ensure that the sub keys are distinct and its precise value has no cryptographic significance.

- The round function differs slightly from a classical Fiestel cipher structure in that integer addition modulo $2^{32}$ is used instead of exclusive-or as the combining operator.

The cipher text as output is then embedded in a video file using Input/Output packages of Java [4].

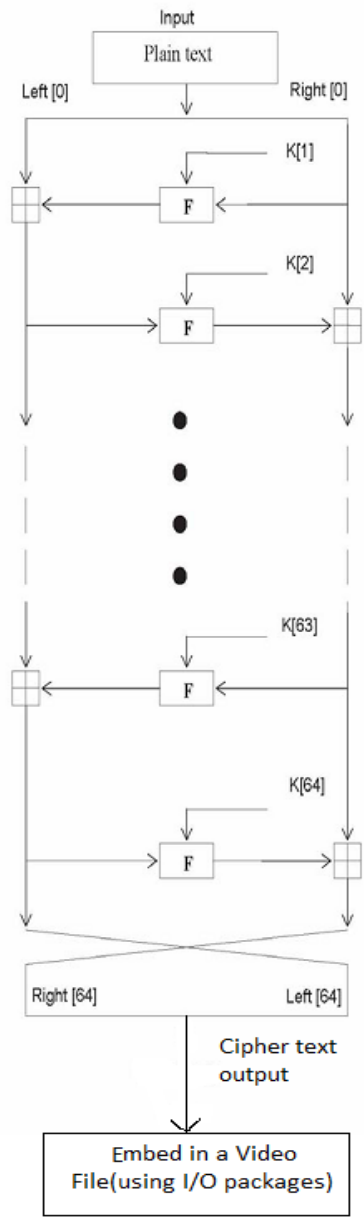

**Fig. 1: Encryption and Embedding**

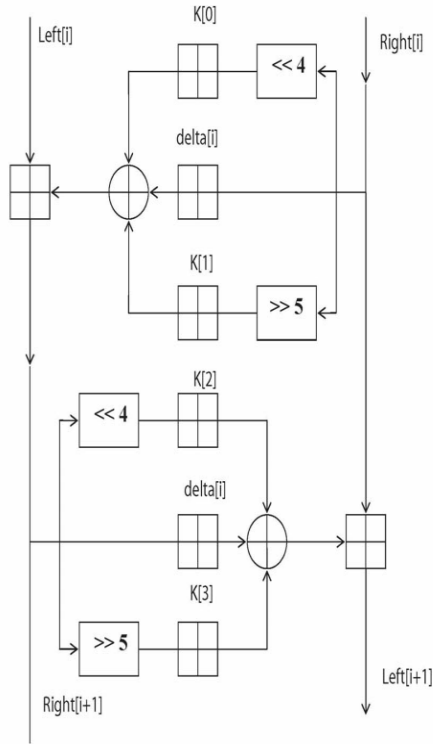

**Fig. 2: Round Function**

Above Figure presents the internal details of the ith cycle of ETEA. The round function, F, consists of the key addition, bitwise XOR and left and right shift operation. We can describe the output (Left [i +1] , Right[i +1] ) of the ith cycle of ETEA with the input (Left[i] ,Right[i] ) as follows

Left [i+1] = Left[i] F (Right[i], K [0, 1], delta[i] ),

Right [i +1] = Right[i] F (Right[i +1], K [2, 3], delta[i] ),

delta[i] = (i +1)/2 * delta,

The round function, F, is defined by

F(M, K[j,k], delta[i] ) = ((M << 4) K[j]) ⊕ (M delta[i] ) ⊕ ((M >> 5) K[k]).

The key schedule algorithm is simple; the 128-bit key K is split into four 32-bit blocks K = (K[0], K[1], K[2], K[3]). The keys K[0] and K[1] are used in the odd rounds and the keys K[2] and K[3] are used in even rounds.

The Embedded message is De-embedded from video file and then Cipher text is taken as input to Decryption process.

Decryption is essentially the same as the encryption process; in the decode routine the cipher text is used as input to the algorithm, but the sub keys K[i] are used in the reverse order.

The intermediate value of the decryption process is equal to the corresponding value of the encryption process with the two halves of the value swapped.

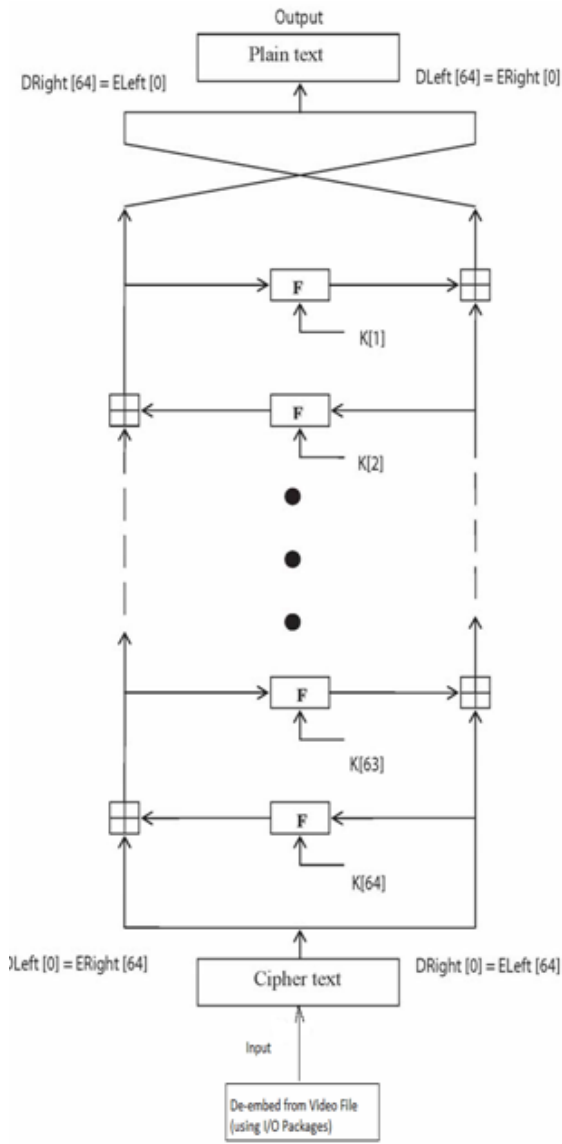

**Fig. 3: De-embedding and Decryption**

## 3. COMPARISON WITH OTHER ALGORITHMS

**Table 1. Comparison between encryption algorithms**

| S.No | Properties | Enhanced Tiny Encryption Algorithm (ETEA) | Data Encryption Standard (DES) | Rivest-Shamir-Adleman (RSA) | Advance Encryption Standard (AES) |
|---|---|---|---|---|---|
| 1. | Key Size | 128 bit Key | 56 bit Key | 1,024 to 4,096 bit typical | 128, 192 or 256 bits |
| 2. | No. of rounds | 64 (32 Cycles) | 16 | 1 | 10, 12 or 14 (depending on key size) |
| 3. | Block Size | 64bits | 64 bits | No Block | 128 bits |
| 4. | Structure | Feistel network | Balanced Feistel network | Structureless | Substitution permutation network |
| 5 | Advantag-e | Provides both encryption, embedding and secure data transmission. | Hardware implementations of DES are very fast. | The biggest advantage of RSA is that it uses asymmetric keys. | Safe Brute Force (128 Bit = 2128 attempts) Unbreakable as for now. |
| 6. | Disadvant-age | ETEA suffers from equivalent keys and can be broken using a related-key attack requiring 223 chosen plaintexts and a time complexity of 232. | DES is now considered insecure because a brute force attack is possible and DES was not designed for software and hence runs slowly. | The prime factors must be kept secret. Anyone can use the public key to encrypt a message, but with currently published methods, if the public key is large enough, only someone with knowledge of the prime factors can feasibly decode the message. | AES-128, the key can be recovered with a computational complexity of 2126.1 using bicliques. |

## 4. CONCLUSION AND FUTURE WORK

We studied TEA, DES, RES and AES and our proposed algorithm supports both the features of cryptography and steganography [6]. So we conclude that although the key size for ETEA is less but it has more number of cycles which makes it better providing the highest security. The major feature of ETEA is that it is both secure and more resistant to attacks.

As ETEA suffers from equivalent keys and can be broken using a related-key attack we include the solution to this problem in our future work.